\documentclass[epj]{webofc}
\pdfoutput=1

\usepackage[utf8]{inputenc}
\usepackage[varg]{txfonts}   
\usepackage{graphicx}
\usepackage{color}
\usepackage{slashed}
\usepackage[english]{babel}
\usepackage[activate={true,nocompatibility}, final, tracking=true, kerning=true, spacing=true, factor=1100, stretch=10, shrink=10]{microtype}
\usepackage{amsmath,amssymb}

\definecolor{ucolor}{rgb}{0.0,0.0,0.3} 
\definecolor{tcolor}{rgb}{0.0,0.0,0.3} 

\usepackage[colorlinks=true,urlcolor=ucolor,anchorcolor=blue,citecolor=blue,filecolor=blue,linkcolor=blue,menucolor=blue,pagecolor=blue,linktocpage=true,pdfa=true]{hyperref}

\newenvironment{eq}{\begin{eqnarray}}{\end{eqnarray}}

\def\tcolor{\color{tcolor}}

\begin{document}
\title{\tcolor Neutrino oscillations in the galactic dark matter halo}
\author{Roberto A. Lineros\thanks{\href{http://goo.gl/00TnL}{INSPIRE-HEP profile}, \email{rlineros@gmail.com}}}

\abstract{
The observation of PeV neutrinos is an open window to study New Physics processes.
Among all possible neutrino observables, the neutrino flavor composition can reveal underlying interactions during the neutrino propagation.
We study the effects on neutrino oscillations of dark matter-neutrino interactions.
We estimate the size of the interaction strength to produce a sizable deviation with respect to the flavor composition from oscillations in vacuum.
We found that the dark matter distribution produces flavor compositions non reproducible by other New Physics phenomena.
Besides, the dark matter effect predicts flavor compositions which depend on the neutrino's arrival direction.
This feature might be observed in neutrino telescopes like IceCube and KM3NET with access to different sky sections.
This effect presents a novel way to test Dark Matter particle models.
}

\maketitle

\section{Introduction}
\label{sec:intro}

The observation and confirmation of neutrino oscillations is one of the earliest evidences of New Physics (NP) beyond the Standard Model (SM) of particle physics.
Roughly speaking, neutrino flavors ($\nu_e$, $\nu_\mu$, and $\nu_\tau$) mix while they propagate through space.
The misalignment between flavor and mass eigenstates is the cause of the oscillations.

The overwhelming observational evidence of neutrino oscillations challenges the SM because it is a clear indication that neutrinos are massive.
A conservative estimation of neutrino masses is $\sum_{i=1}^3 m_{\nu} \lesssim 0.72~{\rm eV}$ which comes from cosmological observations and the analysis done by the PLANCK Collaboration~\cite{Ade:2015xua}.
Moreover, oscillation experiments are sensitive to the neutrino square-mass differences $(\Delta m_{ij}^2 = m_j^2 - m_i^2)$, the mixing angles between flavor/mass eigenstates $(\theta_{ij})$, and the CP-phase $\delta$ (for details check~\cite{Forero:2014bxa} and references within).
Currently, the precision in the neutrino parameters determination allows us to probe possible NP effects.

Neutrinos are unique among other SM particles.
They are the lightest fermions and electrically neutral.
Besides, neutrinos can interact with SM particles via the exchange of $Z$ and $W^{\pm}$ bosons.
This gives them the property to cross through anything without almost interactions.
Therefore neutrinos can unveil regions of the Universe which are hidden to other type of observations.
In this case,  neutrino telescopes like \mbox{IceCube~\cite{Aartsen:2016oji}}, ANTARES~\cite{Michael:2015gjn}, and KM3Net~\cite{Margiotta:2014gza}, among others, give complementary information with respect to other astroparticle messengers (e.g. gamma- and cosmic-rays) becoming unique tools to understand the Universe.
It is important to highlight the IceCube observatory and its leading role on revealing the Universe with the observation of PeV neutrinos~\cite{Aartsen:2015trq}.
This opens the door to test New Physics effects in the neutrino physics sector.
Recently, IceCube has presented results on the flavor composition of astrophysical neutrinos~\cite{Aartsen:2015zva}.
The reported best fit point in the flavor space is far from the expected flavor composition.
The expected composition assumes neutrino oscillations in vacuum and initial neutrino flavor from astrophysics-motivated sources.
In some scenarios, the deviation could be related to the presence of exotic sources of PeV neutrinos~\cite{Bustamante:2015waa}.
Alternatively to such hypothesis, the deviation could come from effects from the presence of Dark Matter (DM).
It is quite well established that the Universe is fulfilled with DM and its presence has a large effect in the cosmological evolution of the Universe.
In fact, the DM relic abundance is quite well estimated using observables extracted from anisotropies of the Cosmic Microwave Background, structure formation, among others.
The combination of all these observables sets the DM relic abundance to $\Omega_{\rm DM} h^2 = 0.1198 \pm 0.0015$~\cite{Ade:2015xua}.

The presence of DM is \emph{per se} an evidence of NP and thus any DM candidate must come from models beyond-SM.
In some cases, the DM candidate is related to neutrino physics (e.g.~\cite{Lattanzi:2014mia} and references within). 
Regardless of the DM nature, it is tantalizing to consider that DM could affect some neutrino observables.
In this manuscript, we summarized the results presented in~\cite{deSalas:2016svi} which studies the neutrino flavor composition of neutrinos propagating in the galactic DM halo.

\section{Neutrino oscillations through a DM medium}
\label{sec:nuosc}

The Milky Way galaxy is embedded in a DM halo. 
The precise DM distribution ($\rho_{\rm DM}$) inside the Milky Way is still under study.
However, it is expected that its DM distribution is in the ballpark of a NFW~\cite{1996ApJ...462..563N} or an isothermal profile distribution~\cite{1980ApJS...44...73B}.

Neutrinos produced inside the Milky Way must travel across the galactic DM halo.
In this scenario, the DM effects on the neutrino oscillation can be modeled using the Mikheyev-Smirnov-Wolfenstein (MSW) effect~\cite{Wolfenstein:1977ue,Mikheev:1986gs}.
In such case the total hamiltonian in the flavor basis is 
\begin{eq}
\mathcal{H}_{\rm tot}(x) = \mathcal{H}_{\rm vac} + \mathcal{V}(x) \, ,
\end{eq}
where $\mathcal{H}_{\rm vac}$ is the (global phase subtracted) hamiltonian in vacuum,
\begin{eq}
\mathcal{H}_{\rm vac} = \frac{1}{2E} U_0 
\left[ \begin{array}{ccc}
0 & 0 & 0 \\
0 & \Delta m^2_{21} & 0 \\
0 & 0 & \Delta m^2_{31} 
\end{array} \right] U_0^{\dagger} \, ,
\end{eq}
which depends on the neutrino square-mass difference $\Delta m^2_{ij} = m^2_i - m^2_j$, the lepton mixing matrix $U_0$~\cite{Agashe:2014kda}, and the neutrino energy $E$.
The spatial dependence of the hamiltonian comes from the effective potential $\mathcal{V}(x)$ in the flavor basis,
\begin{eq}
	\label{eq:pot1}
	\mathcal{V}(x) = \mathcal{V_{\oplus}} \, \hat{\rho}_{\rm DM}(x) \, ,
\end{eq}
where $\hat{\rho}_{\rm DM}(x)$ is the DM distribution (Sun's position normalized) and $\mathcal{V_{\oplus}}$ is the value of the effective potential matrix at the Sun's position.
Depending of the interpretation, the effective potential can be also written mimicking the weak interaction MSW effective potential: 
\begin{eq}
	\label{eq:pot2}
	\mathcal{V}(x) = {\lambda'} \, G_F' N_{\rm DM}(x) \, ,
\end{eq}
where $\lambda'$ is the couplings matrix in the flavor basis, $G_F'= G_F \, m_Z^2/m_{Z'}^2$ is a scaled Fermi constant for a mediator with mass $m_{Z'}$, and \mbox{$N_{\rm DM}(x) = \rho_{\rm DM}(x)/m_{\rm DM}$} is the number density of DM particles.
The relation between Eqs.~\ref{eq:pot1} and \ref{eq:pot2} gives that: 
\begin{eq}
	\label{eq:Vearth}
	\mathcal{V}_{\oplus} = \frac{{\lambda'} \, G_F' \, \rho_{\rm DM}(x_{\oplus})}{m_{\rm DM}} \, ,
\end{eq}
where we use $\rho_{\rm DM}(x_\oplus) = 0.4 \, {\rm GeV}/{\rm cm^3}$ as a common value for all the DM distributions.
Each description of $\mathcal{V}(x)$ remarks different aspects of the effect: Eq.~\ref{eq:pot1} remarks the strength of the effective potential and spatial effects and Eq.~\ref{eq:pot2} remarks a possible particle model extension.
In both cases, the results are the equivalent and provide interesting insights.\\

The evolution of neutrino states in a medium is controlled by
\begin{eq} \label{eq:timeevol}
	i \partial_t \Psi = \mathcal{H}_{\rm tot} \Psi \, ,
\end{eq}
where $\Psi$ is the neutrino field.
The relation between flavor states and mass states is given by
\begin{eq}
	|\nu_{\alpha} \rangle = \sum_{k} U^{*}_{\alpha k} | \nu_{k} \rangle \, ,
\end{eq}
where the greek index $\alpha$ corresponds to a flavor index (e, $\mu$, $\tau$), the latin index $k$ refers to $\mathcal{H}_{\rm tot}$ eigenstates with effective mass $m_{k,{\rm eff}}$, and $U$ is the unitary matrix which diagonalizes $\mathcal{H}_{\rm tot}$:
\begin{eq}
	\mathcal{H}_{\mathrm{tot}}^m  = U^\dagger \mathcal{H}_{\mathrm{tot}} U = \frac{1}{2 E}
	\left( \begin{array}{ccc}
	0 & 0 & 0 \\
	0 & \Delta m_{21,\mathrm{eff}}^2 & 0 \\
	0 & 0 & \Delta m_{31,\mathrm{eff}}^2
	\end{array} \right) \, ,
\end{eq}
where $\mathcal{H}_{\mathrm{tot}}^m$ is the (global phase subtracted) total hamiltonian in the mass-state basis.

The typical distance between sources and Earth ranges from parsecs to kiloparsecs.
This distance is extremely large with respect to oscillation length at the PeV energy making neutrinos to loose coherence.
This allow us to simplify the temporal/spatial evolution of neutrino states because of the large distance averaging effect of neutrino oscillations.
In the case of a spatially homogeneous potential $\mathcal{V}(x) = \mathcal{V}$, the observed flavor composition is:
\begin{eq}
	f_{\beta} = \sum_{\alpha=e,\mu,\tau} \left(\sum_{i=1,3} \left|U_{\beta i} U_{\alpha i}^{*}\right|^2 f^{0}_{\alpha} \right) \, ,
\end{eq}
where $f^0_{\alpha} = (f^0_e,f^0_\mu,f^0_\tau)$ is the flavor composition at the source point, and $U$ is the unitary matrix that diagonalizes the total hamiltonian $\mathcal{H}_{\mathrm{tot}}$.

A more interesting case occurs when the effective potential varies along the neutrino path to Earth.
In this case, we may need to solve Eq.~\ref{eq:timeevol} and follow the evolution of $f_{\beta}$ in different stages along the neutrino path.
However, the DM profile does not vary fast enough and the evolution of neutrino flavors satisfies the adiabatic condition (see~\cite{deSalas:2016svi} for details).
In this case, the flavor composition at Earth is 
\begin{eq}
	f^{\oplus}_{\beta} = \sum_{\alpha=e,\mu,\tau} \left(\sum_{i=1,3} \left|U^{\oplus}_{\beta i} U_{\alpha i}^{0*}\right|^2 f^{0}_{\alpha} \right) \, ,
\end{eq}
where $U^{\oplus}$ diagonalizes $\mathcal{H}_{\rm tot}$ evaluated at Earth's position, while $U^{0}$ diagonalizes $\mathcal{H}_{\rm tot}$ evaluated at the source.
The difference between $U$'s arises because the effective potential scales according to $\rho_{\rm DM}$. 
Let just remark that the matrix texture would not to change during the neutrino path.

\section{Results and Discussion}
\label{sec:results}

The DM effect on the neutrino oscillations depends on the size of $\mathcal{V}_\oplus$ and its texture.
We know from the standard MSW effect that if the effective potential is smaller than $\mathcal{H}^{\rm vac}$ the flavor composition converges to the solution in vacuum.
On the other hand, if it is larger than $\mathcal{H}^{\rm vac}$, the composition is equal to the composition at the source.
Therefore, a rough estimation is $\mathcal{V} \sim \Delta m_{ij}^2/2E$ at some point during the neutrino path.
For neutrinos at PeV range, the effective potential in the range $\mathcal{V} \sim 10^{-17}$~--~$10^{-21}$~eV may affect non-trivially the observed flavor composition.
More details about the exact range and texture of $\mathcal{V}$ are in~\cite{deSalas:2016svi}.

\begin{figure}[t]
	\centering
	\includegraphics[width=0.49\textwidth]{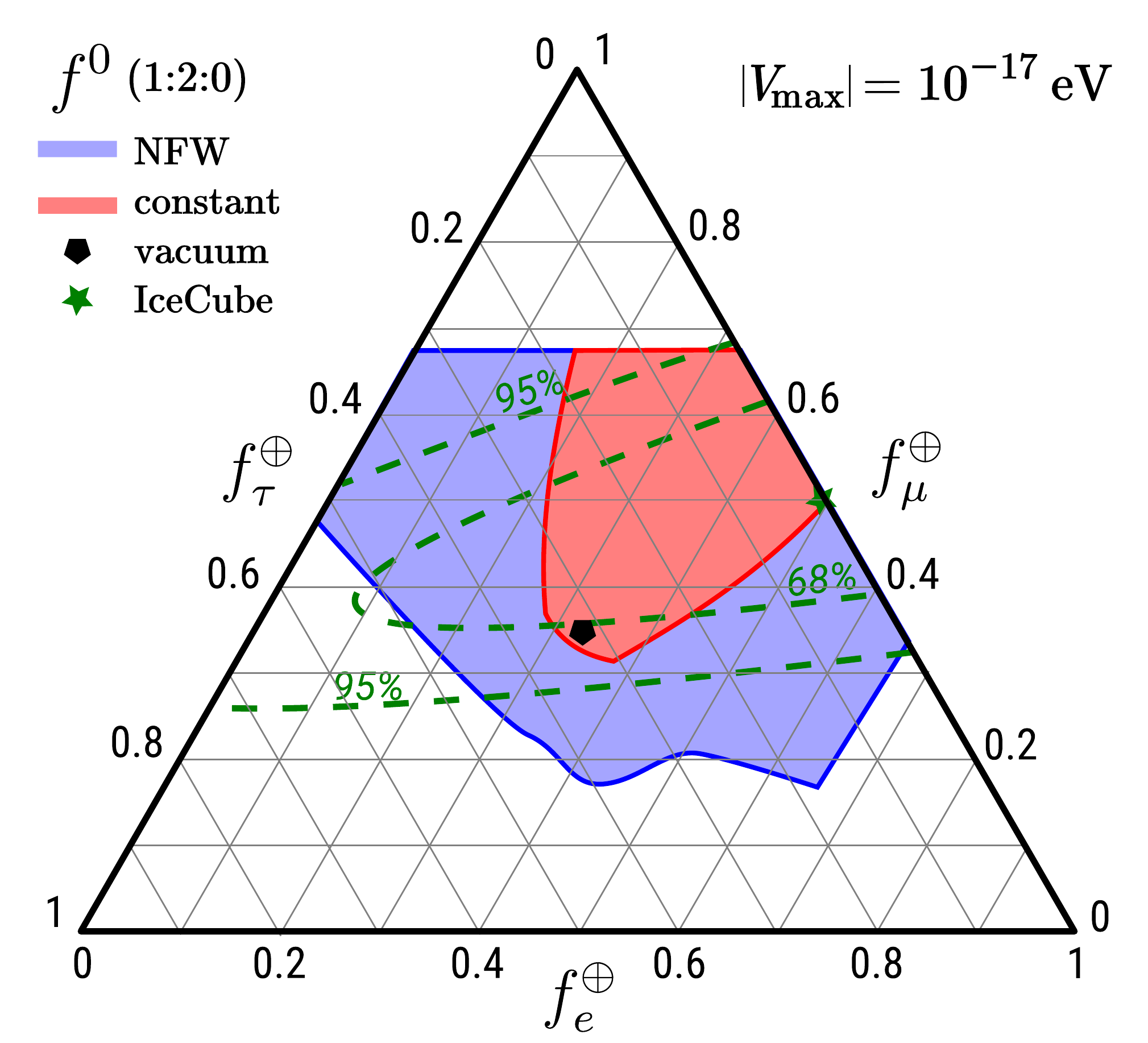} \includegraphics[width=0.49\textwidth]{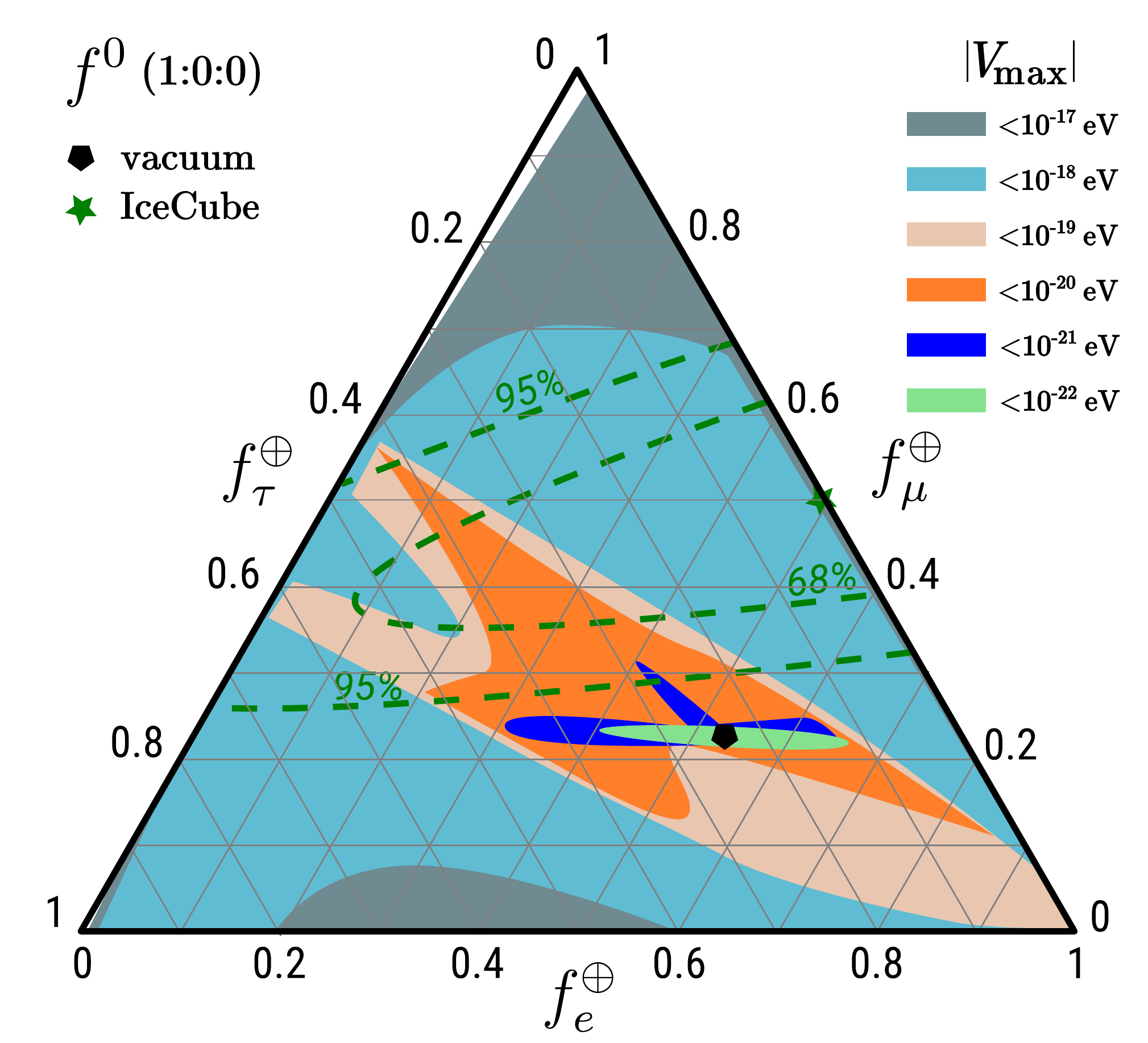}
	\caption{Ternary plots for the neutrino flavor composition at Earth. Left panel shows the accessible flavor regions comparing homogeneous and a NFW DM profile. Right panel shows how the size of the effective potential affects the accessible regions assuming a NFW profile. More details are in~\cite{deSalas:2016svi}.}
	\label{fig:1}
\end{figure}

In Fig.~\ref{fig:1}, we present 2 ternary plots in the flavor space.
In both plots, we consider neutrinos with $E_\nu = 1$~PeV homogeneously produced in the galatic disk.
In the left panel, we assume an initial flavor composition of $f^0 = (1:2:0)$ which is compatible with neutrinos coming from pion decays.
Besides, we randomly generate the matrix $\mathcal{V}_{\oplus}$ keeping $\left|\mathcal{V}_{\oplus}\right| < 10^{-17}$~eV.
The black pentagon indicates the expected composition assuming oscillations in vacuum: $f^{\oplus,{\rm vac}} \simeq (1:1:1)$.
The red region corresponds to the accessible area if the DM distribution is constant.
We highlight that our region coincides with the one coming from Lorentz-violation effects~\cite{Arguelles:2015dca}.
The blue region corresponds to same setup but using a NFW profile.
In this case, the accessible region for the NFW profile is larger than the constant profile case.
This indicates that spatial dependence of the effective potential does largely affect the observed flavor composition.
The green star indicates IceCube's best fit.
Unfortunately, its location is just in the interface among both areas.

The right panel of same figure shows the accessible region when maximum value of $\mathcal{V}$ is restricted.
This case uses $f^0 = (1:0:0)$ and a NFW profile.
The black pentagon is the expected flavor composition in vacuum and it corresponds to $f^{\oplus,{\rm vac}} \simeq (0.56:0.22:0.22)$.
Here, we show that the effect of the DM distribution can cover almost the full flavor triangle for $\mathcal{V}$ values as large as $10^{-17}$~eV.
Smaller values of the potential shrinks the area, converging to an small area around the vacuum solution. \\

The $\mathcal{V}$ scale may indicate a beyond-SM solution.
To this end, we consider DM models like asymmetric-DM, scalar DM, and fuzzy-DM~\cite{Hu:2000ke}.
All of them could generate such scale providing interesting classes of models.
For the interpretation in terms of particle models, we need to calculate the neutrino mean free path,
\begin{eq}
	l_{\nu} = \left( \sigma \, N_{\rm DM} \right)^{-1} \, ,
\end{eq}
which indicates the distance that neutrinos can cover without (hard) scattering on DM.
For distances $\mathcal{O}\left( {\rm kpc} \right)$, i.e. Milky Way size, we required roughly $\sigma \lesssim \times 10^{-21} \, {\rm cm}^2 \times m_{\rm DM} / {\rm GeV}$.
This condition ensures that the DM interaction only affects the neutrino oscillation. 

The mean free path depends on same parameters as Eq.~\ref{eq:Vearth}.
The cross section for a $t$-channel interaction in the high energy regime ($\sqrt{s} \gg m_Z',m_{\rm DM}$) becomes
\begin{eq}
	\sigma \simeq \lambda' \left( \frac{m_{Z'}}{\rm GeV} \right)^{-2} \times 3.75 \times 10^{-29} {\rm cm}^2 \, .
\end{eq}
Looking inside Eq.~\ref{eq:Vearth}, we can estimate the numerical value of two parameters among $\lambda'$, $m_{\rm DM}$ and $G_F'$ (or $m_{Z'}$) by reproducing the required value of $\mathcal{V}$ and $l_{\nu}$.
We found that fuzzy-DM models with $m_{\rm DM} \simeq 10^{-23}$~eV align better if the interaction strength is similar to the one of Weak interactions.
Models with masses in the GeV--keV range, inspired by WIMPs, majorons, etc. (e.g.~\cite{Hirsch:2013ola, Lineros:2014jba, Rojas:2017sih}), requires mediators with masses $m_Z'\sim 10^{-6} (10^{-2})$~eV which would generate an observable effect in the neutrino oscillation.
We study other possibilities including making the mean free path larger than the observable universe (see details in~\cite{deSalas:2016svi}).\\

Another interesting aspect of the DM effects is the dependence on the arrival direction. 
Galactic neutrinos originated in regions closer to the galactic center would be affected differently that those originated in the Milky Way's outskirts.
This is important because neutrino telescopes like IceCube and KM3NET are sensitive to different patches of the sky.
Indeed, this effect indicates that observed flavor composition in both telescopes may be different.
The case of extragalactic neutrinos are also interesting because it allows us to explore DM densities in extragalactic halos.

\section{Conclusions}
\label{sec:concl}

The observation of PeV neutrinos allows us to study the Universe as never done before.
Similarly, the presence of DM in the Universe is a clear signal of physics beyond-SM.

Neutrinos traveling to Earth must propagate through the galactic DM halo.
Assuming that DM and neutrinos do interact, we study the effect of such interactions at the level of neutrino oscillations.
We found that the effect can produce flavor compositions at Earth different than those expected from neutrino oscillations in vacuum.
Besides, the spatial dependence generates flavor compositions outside the region obtained from other NP effects.

The interpretation in terms of DM models gives interesting constructions.
For DM particles with masses in the keV and GeV range, the neutrino-DM interaction could be mediated by new particles with masses in the sub-eV range.
The implication for the Weak-like interaction is that DM mass has to be extremely small $\mathcal{O}(10^{-23} \, {\rm eV})$.
This solution is compatible with fuzzy-DM models.

Better statistics of the neutrino flavor composition might indicate a dependence with respect to the neutrino arrival direction.
This feature gives a nice prediction of this effect.\\

\begin{acknowledgement}
I would like to thank the RICAP 2016 organizers for a very interesting conference.
I thank \href{http://goo.gl/CHlgQR}{P.~F.~de~Salas} and \href{http://goo.gl/x8CPgJ}{M.~T\'{o}rtola} for their great work done in the paper referred in this manuscript.
I also thank the hospitality of the Theoretical Physics group at the U. Libre de Bruxelles and the \emph{Interactions Fondamentales en Physique et en Astrophysique} (IFPA) group at U. Li\`ege.
\href{http://goo.gl/00TnL}{R.~A.~L.} acknowledges the support of the Juan de la Cierva contract JCI-2012-12901 (MINECO) and the Spanish MESS via the {\it Servicio P\'ublico de Empleo Estatal}.
\end{acknowledgement}


\def\apj{Astrophys.~J.}                       
\def\apjl{Astrophys.~J.~Lett.}                
\def\apjs{Astrophys.~J.~Suppl.~Ser.}          
\def\aap{Astron.~\&~Astrophys.}               
\def\aj{Astron.~J.}                           %
\def\araa{Ann.~Rev.~Astron.~Astrophys.}       %
\def\mnras{Mon.~Not.~R.~Astron.~Soc.}         %
\def\physrep{Phys.~Rept.}                     %
\def\jcap{J.~Cosmology~Astropart.~Phys.}      
\def\jhep{J.~High~Ener.~Phys.}                
\def\prl{Phys.~Rev.~Lett.}                    
\def\prd{Phys.~Rev.~D}                        
\def\nphysa{Nucl.~Phys.~A}                    

\bibliography{biblio.bib}

\end{document}